# 2D Riemann-Christoffel curvature tensor via a 3D space using a specialized permutation scheme

Mensur Omerbashich
Department of Geodesy & Geomatics Engineering, University of New Brunswick, Fredericton, Canada
(First published in 2003, reference as: *Geodetski Glasnik BiH* 37:5-12 (2003))
AMS CLASSIFICATION NUMBERS: 53B21, 86-08

Abstract: When a space in which Christoffel symbols of the second kind are symmetrical in lower indices exists, it makes for a supplement to the standard procedure ($\varphi, \lambda, h$)→($\theta, \lambda, r$) when a 2D surface is normally induced from the geometry of the surrounding 3D space in which the surface is embedded. There it appears appropriate to use a scheme for straightforward permutation of indices of $\Gamma^k_{ij}$, when such a space would make this transformation possible, so as to obtain the components of the 2D Riemann-Christoffel tensor (here expressed in geodetic coordinates for an ellipsoid of revolution, of use in geophysics). By applying my scheme I find the corresponding indices in 2D and 3D supplement-spaces, and I compute components of the Riemann-Christoffel tensor. By operating over the elements of the projections alone, the all-known value of $1/MN$ for the Gaussian curvature on an ellipsoid of revolution is obtained. To further validate my scheme, I show that in such a 3D space the tangent vector to a $\phi$-curve for $\lambda=\text{const}_1$ would be parallel to a tangent vector to a $\phi$-curve for $\lambda=\text{const}_2$ on the surface of an ellipsoid of revolution. Surfaces parameterized by Gauss surface normal coordinates ($\varphi_{el}, \lambda_{el}, h_{el}(\varphi_{el}, \lambda_{el})$), such as the Earth, now can have the Riemann-Christopher curvature tensor computed in a straightforward fashion for the topographic surface $h_{el}(\varphi_{el}, \lambda_{el})$ of the Earth, given Christoffel symbols for such a representation in terms of orthonormal functions on the ellipsoid of revolution.

Introduction

I derive the metric tensor for geodetic coordinates ($\varphi, \lambda, h$) on a fixed ellipsoid of revolution given by its two axes ($a, b$), and centered upon the coordinate origin for which $x^1 = (N+h)\cos\varphi\cos\lambda$, $x^2 = (N+h)\cos\varphi\cos\lambda$, $x^3 = (N+h)\cos\varphi\cos\lambda$, as [Hotine, 1969]:

$$g_{ps} = \delta_{ij} \frac{\partial x^i}{\partial u^p} \frac{\partial x^j}{\partial u^s} \qquad (1)$$

where $N = a\sqrt{\cos^2\varphi + (b/a)^2 \sin^2\varphi}$, and the $u^i \in U$ are the curvilinear coordinates in a curvilinear coordinate system U, e.g., $\varphi, \lambda, h$ on an ellipsoid of revolution. I then obtain from (1):

$$\left.\begin{aligned}
g_{11} &= \cos^2\varphi\cos^2\lambda + \cos^2\varphi\sin^2\lambda + \sin^2\varphi = 1 \\
g_{22} &= \frac{b^4}{a^8}N^6\sin^2\varphi\cos^2\lambda + \frac{b^4}{a^8}N^6\sin^2\varphi\sin^2\lambda + \frac{b^4}{a^8}N^6\cos^2\varphi = \frac{b^4}{a^8}N^6 \\
g_{33} &= (N+h)^2\cos^2\varphi\sin^2\lambda + (N+h)^2\cos^2\varphi\cos^2\lambda + 0 = (N+h)^2\cos^2\varphi, \\
&\text{or for } h=0: \quad \ldots = N^2\cos^2\varphi = p^2 \quad \forall\, p \neq s \Rightarrow g_{ps} = 0
\end{aligned}\right\} \qquad (2)$$

and the metric:

$$g_{ij} = \begin{Vmatrix} 1 & 0 & 0 \\ 0 & (M+h)^2 & 0 \\ 0 & 0 & (N+h)^2\cos^2\varphi \end{Vmatrix}. \qquad (3)$$

Christoffel symbols of the 1st and the 2nd kind are respectively given by [Wrede, 1963]:

$$[ij, k] = \frac{1}{2}\left(\frac{\partial g_{jk}}{\partial u^i} + \frac{\partial g_{ik}}{\partial u^j} - \frac{\partial g_{ij}}{\partial u^k}\right), \qquad \Gamma_{ij}^k = g^{lk} \cdot [ij, k], \qquad (4)$$

which for an ellipsoid of revolution in geodetic coordinates $(x^1, x^2, x^3) \equiv (h, \varphi, \lambda)$ read:

$$\left.\begin{aligned}
\Gamma_{23}^3 &= \Gamma_{32}^3 = g^{33} \cdot [32,3] = \frac{1}{N^2 \cos^2 \varphi} \cdot (-MN \sin\varphi \cos\varphi) = -\frac{M}{N}\tan\varphi \\
\Gamma_{13}^3 &= \Gamma_{31}^3 = g^{33} \cdot [13,3] = \frac{1}{N^2 \cos^2 \varphi} \cdot N \cos^2\varphi = \frac{1}{N} \\
\Gamma_{12}^2 &= \Gamma_{21}^2 = g^{22} \cdot [12,2] = \frac{1}{M^2} \cdot M = \frac{1}{M} \\
\Gamma_{22}^2 &= g^{22} \cdot [22,2] = \frac{1}{M^2} \cdot 3M^2\left(1 - \frac{M}{N}\right)\tan\varphi = 3\cdot\left(1 - \frac{M}{N}\right)\tan\varphi \\
\Gamma_{22}^1 &= g^{11} \cdot [22,1] = \frac{1}{1} \cdot (-M) = -M \\
\Gamma_{33}^1 &= g^{11} \cdot [33,1] = \frac{1}{1} \cdot (-N\cos^2\varphi) = -N\cos^2\varphi \\
\Gamma_{33}^2 &= g^{22} \cdot [33,2] = \frac{1}{M^2} \cdot MN\sin\varphi\cos\varphi = \frac{N}{M}\sin\varphi\cos\varphi.
\end{aligned}\right\} \qquad (5)$$

Due to generality of tensors, the sufficient proof of appropriateness of such a space for considerations within the Euclidean space is made by subjecting it to the same constraint. Namely, there, the vector field obtained by parallel transport of a surface vector along a curve – geodesic (a straight line in that case), and given by [Spiegel, 1959]:

$$\frac{\partial A^r}{\partial x^k} + \Gamma_{jk}^r A^j = 0, \qquad (6)$$

cuts equal angles with the geodesic (straight line) as it itself traces out one straight line.

  The concept of parallelism in the Riemannian space is defined relative to a given curve. On an ellipsoid of revolution – being one such space – there is always a geodesic that connects two given points. Thus, for tangent vectors to a φ-curve for $\lambda = \lambda_1$ =const., and to a φ-curve for $\lambda = \lambda_2$ = const., regardless of the existence of parallelism in the Euclidean sense, we speak of parallelism if parallel transport between those two tangent vectors occurred, i.e., if the above eqn. (6) is satisfied. This clearly is the case in the Euclidean space where all the Christoffel symbols vanish, and where all the remaining derivatives of straight lines are equal to zero: meridians are parallel in transverse mapping projections.

  In order to see if the said tangent vectors are generally parallel on an ellipsoid of revolution, it suffices to show that at least one of the system of equations (6) does not satisfy for (5):

$$\frac{\partial A^3}{\partial x^3} + \Gamma_{23}^3 \cdot A^2 = \frac{\partial}{\partial \lambda}(0) - \frac{M}{N}\tan\varphi \cdot M^2 = -\frac{M^3}{N}\tan\varphi \neq 0 \text{ except for } \varphi = 0 \wedge \varphi = 2\pi. \quad (7)$$

Derivation of the Riemann-Christoffel tensor

The Riemann-Christoffel (curvature) tensor reads [Wrede, 1963]:

$$R^l_{ijk} = \frac{\partial \Gamma^l_{ik}}{\partial u^j} - \frac{\partial \Gamma^l_{ij}}{\partial u^k} + \Gamma^s_{ik}\Gamma^l_{sj} - \Gamma^s_{ij}\Gamma^l_{sk}, , \qquad (8)$$

while its covariant form is defined as:

$$R_{lijk} = g_{ls}R^{ks}_{ijk}. \qquad (9)$$

Its only four non-zero components in a 2D symmetrical space are $R_{1212}$, $R_{2121}$, $R_{1221}$, $R_{2112}$, where due to symmetry in the two pairs of indices, as well as to anti-symmetry, it is also $R_{1212} = R_{2121} = -R_{1221} = -R_{2112}$ [Wrede, 1963].

    A sphere and an ellipsoid of revolution represent 2D curved spaces (surfaces). For Christoffel symbols (Appendix), I used 3D coordinates. Therefore, I will now use for computing the curvature tensor the following scheme for permutation of indices of Christoffel symbols of the 2$^{nd}$ kind and of corresponding elements of the Riemann-Christoffel tensor(s):

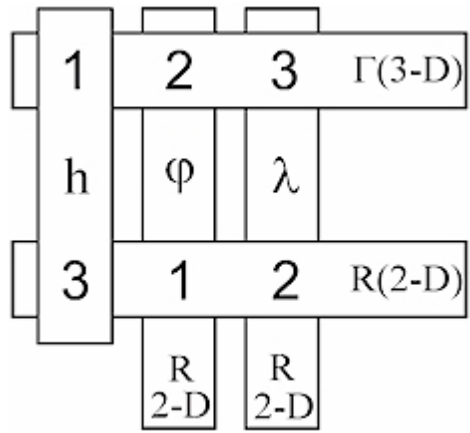

    The above scheme is used so that its columns and rows are cross-referenced to find out about the indices corresponding to each other in a 2D and 3D space. For an ellipsoid of revolution, the components of the Riemann-Christoffel curvature tensor, after (8) and applying the above scheme, now read:

$$R^1_{212} = \frac{\partial \Gamma^2_{33}}{\partial \varphi} - \frac{\partial \Gamma^2_{32}}{\partial \lambda} + \Gamma^2_{33}\Gamma^2_{22} - \Gamma^3_{32}\Gamma^2_{33} =$$

$$= \frac{\partial}{\partial \varphi}\left(\frac{N}{M}\sin\varphi\cos\varphi\right) - 0 + \frac{N}{M}\sin\varphi\cos\varphi \cdot 3\left(1 - \frac{M}{N}\right)\tan\varphi + \frac{M}{N}\tan\varphi \cdot \frac{N}{M}\sin\varphi\cos\varphi =$$

$$= \frac{\partial}{\partial \varphi}\left(\frac{1}{2}\cdot\frac{N}{M}\sin 2\varphi\right) + 3\frac{N}{M}\sin\varphi\cos\varphi \cdot \frac{\sin\varphi}{\cos\varphi} - 3\frac{N}{M}\cdot\frac{M}{N}\sin\varphi\cos\varphi \cdot \frac{\sin\varphi}{\cos\varphi} + \sin^2\varphi =$$

$$= \frac{1}{2}\left[-2\left(\frac{N}{M} - 1\right)\tan\varphi \cdot 2\sin\varphi\cos\varphi + \frac{N}{M}\cdot 2\cos 2\varphi\right] + 3\frac{N}{M}\sin^2\varphi - 3\sin^2\varphi + \sin^2\varphi =$$

$$= -2\left(\frac{N}{M} - 1\right)\frac{\sin\varphi}{\cos\varphi}\sin\varphi\cos\varphi + \frac{N}{M}\left(\cos^2\varphi - \sin^2\varphi\right) + 3\frac{N}{M}\sin^2\varphi - 2\sin^2\varphi =$$

$$= -2\frac{N}{M}\sin^2\varphi + 2\sin^2\varphi + \frac{N}{M}\cos^2\varphi - \frac{N}{M}\sin^2\varphi + 3\frac{N}{M}\sin^2\varphi - 2\sin^2\varphi =$$

$$= \frac{N}{M}\cos^2\varphi - \sin^2\varphi + \sin^2\varphi \Rightarrow$$

$$R^1_{212} = \frac{N}{M}\cos^2\varphi. \tag{10}$$

Computation of other three components is left to the reader. The covariant components are now computed from (9) when applied to (10):

$$R_{1212} = g_{11} \cdot R^1_{212} = MN\cos^2\varphi, \tag{11}$$

which complies with the discussion immediately following eqn. (9).

For completeness, I now first derive the Ricci-Einstein and Lamé tensors $R_{ij} = g^{mn}R_{mijn}$ and $S^{ij} = e^{ikl}e^{jmn}R_{klmn}$ [Borisenko & Tarapov, 1966] respectively, for an ellipsoid of revolution when (see Appendix):

$$R_{ij} = \left\|\begin{array}{cc} -\dfrac{M}{N} & 0 \\ 0 & -\dfrac{N}{M}\cos^2\varphi \end{array}\right\|. \tag{12}$$

I also obtain:

$$S^{11} = \frac{-1}{\sqrt{g}}\frac{1}{\sqrt{g}}R_{1212} = -\frac{1}{g}R_{1212} = -\frac{MN\cos^2\varphi}{M^2N^2\cos^2\varphi} = -\frac{1}{MN}. \tag{13}$$

where [Wrede, 1963]:

$$e^{ikl} = \frac{1}{\sqrt{g}}\delta_{ikl} \wedge e^{jmn} = \frac{1}{\sqrt{g}}\delta_{jmn}, \text{ and } S^{ij} = S^{ji}. \tag{14}$$

The remaining components are obtained in an analogous way, and are left for the reader to derive.

Applying the formula for Gaussian curvature from Riemann-Christoffel curvature tensor $K = (1/g)\,R_{1212}$ [Sokolnikoff, 1964] to an ellipsoid of revolution finally yields:

$$K_{Elipsoid} = \frac{1}{M^2 N^2 \cos^2 \varphi} \cdot MN \cos^2 \varphi = \frac{1}{MN}. \tag{15}$$

Conclusions

The correctness of (15), seen here as obtained from the above formula for Gaussian curvature, complies with (v) from the Appendix. The scheme enables a straightforward transformation into the Euclidean space, i.e., from an arbitrary (here flat, but easily extended to general case) 3D to the 2D space embedding in the Euclidean sense the ellipsoid of revolution. The next step in theoretical considerations would now be to describe the surfaces parameterized by Gauss surface normal coordinates ($\varphi_{el}$, $\lambda_{el}$, $h_{el}(\varphi_{el}, \lambda_{el})$), such as the Earth. This possible application could see the Riemann-Christopher curvature tensor computed in a straightforward fashion for the topographic surface $h_{el}(\varphi_{el}, \lambda_{el})$ of the Earth [Engels and Grafarend, 1992], given the Christoffel symbols for such a representation in terms of orthonormal functions on the ellipsoid of revolution up to degree and order 180 [Grafarend and Keller, 1995].

Appendix

Some useful expressions:

$$M = N^3 \frac{b^2}{a^4} \;;\; \frac{\partial M}{\partial \varphi} = 3M\left(1 - \frac{M}{N}\right) \tan \varphi$$

$$\frac{\partial}{\partial \varphi}\left(\frac{M}{N}\right) = 2\frac{M}{N} \cdot \left(1 - \frac{M}{N}\right) \tan \varphi \;;\; \frac{\partial}{\partial \varphi}\left(\frac{N}{M}\right) = -2\left(\frac{N}{M} - 1\right) \tan \varphi \tag{i}$$

The expressions (4) were obtained after the next computations took place:

$$\frac{\partial x^1}{\partial \varphi} = -\frac{b^2}{a^4} N^3 \sin \varphi \cos \lambda$$

$$\frac{\partial x^2}{\partial \varphi} = -\frac{b^2}{a^4} N^3 \sin \varphi \sin \lambda \tag{ii}$$

$$\frac{\partial x^3}{\partial \varphi} = \frac{b^2}{a^4} pN^2 = \frac{b^2}{a^4} N^4 \cos^2 \varphi.$$

In the above, I found the complete derivatives of $N = f(\varphi)$. Let it be mentioned here that the metric can be arrived at in several ways; I used the well known way by Hotine [1969]:

$$\frac{\partial}{\partial \varphi}(N \cos \varphi) = -M \sin \varphi \quad \wedge \quad \frac{\partial}{\partial \varphi}(N \sin \varphi) = \frac{N - M}{\cos \varphi} + M \cos \varphi \tag{iii}$$

where (i) could serve as a proof of the correctness of the result obtained in either way.

By evaluating (10), I obtain the non-zero Christoffel symbols of the first kind for a 3D space given in geodetic coordinates, with respect to an ellipsoid $h = 0$:

$$[32,3] = \frac{1}{2}\left[\frac{\partial g_{23}}{\partial x^3} + \frac{\partial g_{33}}{\partial x^2} - \frac{\partial g_{32}}{\partial x^3}\right] = 0 + \frac{1}{2}\frac{\partial}{\partial \varphi}(N^2 \cos^2 \varphi) - 0 = -MN \sin \varphi \cos \varphi$$

$$[13,3] = \frac{1}{2}\left[\frac{\partial g_{33}}{\partial x^1} + \frac{\partial g_{13}}{\partial x^3} - \frac{\partial g_{13}}{\partial x^3}\right] = \frac{1}{2}\frac{\partial}{\partial h}(N^2 \cos^2 \varphi) + 0 - 0 = N \cos^2 \varphi$$

$$[12,2] = \frac{1}{2}\left[\frac{\partial g_{22}}{\partial x^1} + \frac{\partial g_{12}}{\partial x^2} - \frac{\partial g_{12}}{\partial x^2}\right] = \frac{1}{2}\frac{\partial}{\partial h}(M^2) + 0 - 0 = M$$

$$[22,2] = \frac{1}{2}\left[\frac{\partial g_{22}}{\partial x^2} + \frac{\partial g_{22}}{\partial x^2} - \frac{\partial g_{22}}{\partial x^2}\right] = \frac{1}{2}\frac{\partial}{\partial \varphi}(M^2) = 3M^2\left(1 - \frac{M}{N}\right)\tan \varphi$$

$$[22,1] = \frac{1}{2}\left[\frac{\partial g_{21}}{\partial x^2} + \frac{\partial g_{21}}{\partial x^2} - \frac{\partial g_{22}}{\partial x^1}\right] = 0 + 0 - \frac{1}{2}\frac{\partial}{\partial h}(M^2) = -M$$

$$[33,1] = \frac{1}{2}\left[\frac{\partial g_{31}}{\partial x^3} + \frac{\partial g_{31}}{\partial x^3} - \frac{\partial g_{33}}{\partial x^1}\right] = 0 + 0 - \frac{1}{2}\frac{\partial}{\partial h}(N^2 \cos^2 \varphi) = -N \cos^2 \varphi$$

$$[33,2] = \frac{1}{2}\left[\frac{\partial g_{32}}{\partial x^3} + \frac{\partial g_{32}}{\partial x^3} - \frac{\partial g_{33}}{\partial x^2}\right] = 0 + 0 - \frac{1}{2}\frac{\partial}{\partial \varphi}(N^2 \cos^2 \varphi) = MN \sin \varphi \cos \varphi \qquad \text{(iv)}$$

$$[23,3] = \frac{1}{2}\left[\frac{\partial g_{33}}{\partial x^2} + \frac{\partial g_{23}}{\partial x^3} - \frac{\partial g_{23}}{\partial x^3}\right] = \frac{1}{2}\frac{\partial}{\partial \varphi}(N^2 \cos^2 \varphi) + 0 - 0 = -MN \sin \varphi \cos \varphi$$

$$[31,3] = \frac{1}{2}\left[\frac{\partial g_{13}}{\partial x^3} + \frac{\partial g_{33}}{\partial x^1} - \frac{\partial g_{31}}{\partial x^3}\right] = 0 + \frac{1}{2}\frac{\partial}{\partial h}(N^2 \cos^2 \varphi) - 0 = N \cos^2 \varphi$$

$$[21,2] = \frac{1}{2}\left[\frac{\partial g_{12}}{\partial x^2} + \frac{\partial g_{22}}{\partial x^1} - \frac{\partial g_{21}}{\partial x^2}\right] = 0 + \frac{1}{2}\frac{\partial}{\partial h}(M^2) - 0 = M.$$

From (4) it is obvious that only for

$$[ij,k] \neq 0 \Rightarrow \Gamma_{ij}^k \neq 0.$$

The curvature of an ellipsoid is computed using the above expressions as:

$$K_{Ellipsoid} = -\frac{1}{2\sqrt{g}} \cdot \left[\frac{\partial}{\partial \varphi}\left(\frac{1}{\sqrt{g}}\frac{\partial g_{22}}{\partial \varphi}\right) + \frac{\partial}{\partial \lambda}\left(\frac{1}{\sqrt{g}}\frac{\partial g_{11}}{\partial \lambda}\right)\right] =$$

$$= -\frac{1}{2 \cdot \frac{b^2}{a^4}N^3 p} \cdot \left[\frac{\partial}{\partial \varphi}\left(\frac{1}{\frac{b^2}{a^4}N^3 p} \cdot \frac{\partial(p^2)}{\partial \varphi}\right) + \frac{\partial}{\partial \lambda}\left(\frac{1}{\frac{b^2}{a^4}N^3 p} \cdot \frac{\partial\left(\frac{b^2}{a^4}N^3 p\right)}{\partial \lambda}\right)\right] =$$

$$= -\frac{1}{2 \cdot \frac{b^2}{a^4}N^3 p} \cdot \frac{\partial}{\partial \varphi}\left(\frac{2a^4 p}{b^2 N^3 p} \cdot \frac{dp}{d\varphi}\right) = \frac{-a^4}{2b^2 N^3 p} \cdot \frac{2a^4}{b^2} \cdot \frac{\partial}{\partial \varphi}\left(\frac{1}{N^3} \cdot \frac{-b^2}{a^4} N^3 \sin \varphi\right) =$$

$$= \frac{a^8}{b^4 N^3 p} \cdot \frac{b^2}{a^4}\cos \varphi = \frac{a^4}{b^2} \cdot \frac{\cos \varphi}{N^4 \cos \varphi} = \frac{a^4}{b^2 N^4} = \frac{1}{MN}. \qquad \text{(v)}$$

Acknowledgments

I thank Drs. Erik Grafarend and Juraj Janák for helpful comments.